\documentclass[prl,twocolumn,a4paper,superscriptaddress]{revtex4}
\usepackage[english]{babel}
\usepackage{amsmath}
\usepackage{amsfonts}
\usepackage{amssymb}
\usepackage{array} 
\usepackage{dcolumn}
\usepackage{longtable}
\usepackage{hyperref}
\usepackage{float}
\usepackage{bbm}
\usepackage{bm}
\usepackage{graphicx}
\usepackage{natbib}

\begin{document}
\title{Extensile actomyosin?}
\date{\today}
\author{Martin Lenz}
\email{martin.lenz@u-psud.fr}
\affiliation{LPTMS, CNRS, Univ. Paris-Sud, Universit\'e Paris-Saclay, 91405 Orsay, France}

\begin{abstract}
Living cells move thanks to assemblies of actin filaments and myosin motors that range from very organized striated muscle tissue to disordered intracellular bundles. The mechanisms powering these disordered structures are debated, and all models studied so far predict that they are contractile. We reexamine this prediction through a theoretical treatment of the interplay of three well-characterized internal dynamical processes in actomyosin bundles: actin treadmilling, the attachement-detachment dynamics of myosin and that of crosslinking proteins. We show that these processes enable an extensive control of the bundle's active mechanics, including reversals of the filaments' apparent velocities and the possibility of generating extension instead of contraction. These effects offer a new perspective on well-studied \emph{in vivo} systems, as well as a robust criterion to experimentally elucidate the underpinnings of actomyosin activity.
\end{abstract}

\maketitle

\begin{figure}[b]
\includegraphics[width=\columnwidth]{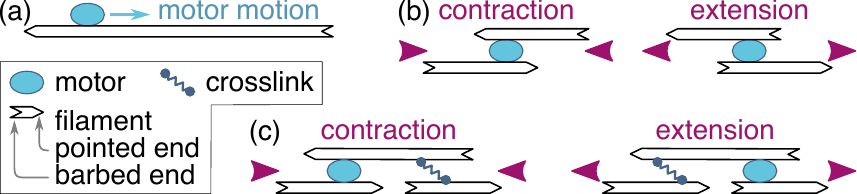}
\caption{\label{fig:mechanism}Actomyosin bundle dynamics involves a competition between contraction and extension. 
(a)~Motors bind filaments and move towards their barbed ends.
(b)~This motion results in local contraction or extension depending on the local arrangement of the filaments.
(c)~In a full bundle, a given filament arrangement can generate contraction or extension depending on the localization of the motors and crosslinks. The present work shows that motor and crosslink self-organization can bring about either outcome.
}
\end{figure}

{\noindent}Many cellular functions, from motility to cell division, are driven by myosin motors exerting forces on actin filaments held together by crosslinking proteins. This wide variety of processes is powered by an equally wide range of actomyosin structures, many of which do not display any apparent spatial organization of their components~\cite{Verkhovsky:1995,Cramer:1997,Medalia:2002,Kamasaki:2007}. While these structures are overwhelmingly observed to contract~\cite{Murrell:2015}, the mechanisms underlying this contraction are unclear, as individual myosin motors can in principle elicit extension just as easily as contraction [Fig.~\ref{fig:mechanism}(a-b)]~\cite{Hatano:1994,Sekimoto:1998,Lenz:2012a,Pinto:2013}.

Recent investigations into this breaking of symmetry between contraction and extension have focused on two classes of models. The first of these classes is based on the idea that \emph{mechanical nonlinearities}, \emph{e.g.}, the buckling of individual filaments under compression could suppress the propagation of extensile forces and thus favor contraction~\cite{Dasanayake:2011,Lenz:2012,Ronceray:2016}. 
By contrast, in the second type of models the spatial \emph{self-organization} of the bundle's motors and crosslinks along undeformable, rod-like actin filaments leads to contraction~\cite{Kruse:2002,Zumdieck:2007,Zemel:2009,Oelz:2015}.
So far, opportunities to discriminate between these two models experimentally remain very limited for lack of a clear theoretical prediction setting one apart from the other.

Here we provide such a prediction, namely that the self-organization mechanisms imply that actomyosin bundles robustly \emph{extend} if taken to certain parameter regimes. This stark qualitative change from contraction to extension is easily detectable experimentally, and is not expected in mechanical nonlinearities models.
Our prediction crucially rests on a simultaneous treatment of the filament, motor and crosslink dynamics; previous studies only involved partial treatments. The coupled dynamics of these elements induces a spatial organization of motors and crosslinks along the filaments, and our predicted switch between contraction and extension is driven by a localization of the motors and crosslinks to the filament ends, as illustrated in Fig.~\ref{fig:mechanism}(c). We characterize the experimental regimes where either behavior is expected, and find that extension arises when the motor run-length and unbinding rate are relatively large compared to the filament length and the crosslink unbinding rate, respectively. Our study moreover identifies simple, widely applicable ideas to understand self-organization in active filament-motor systems.

\begin{figure*}[t]
\centering
\includegraphics[width=\textwidth]{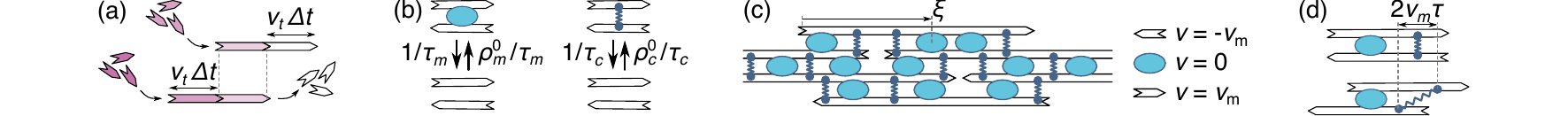}
\caption{\label{fig:parametrization}Principles of filament, motor and cross-link dynamics. (a)~Simultaneous polymerization at the barbed end (incoming purple monomers) and depolymerization from the pointed end (outgoing white monomers) induce a leftwards ``treadmilling'' motion of the filament. The top and bottom images respectively correspond to times $t$ and $t+\Delta t$. (b)~Motors come on and off a pair of filaments with constant rates (on the left), and so do crosslinks (on the right). (c)~In an assembly of identical filaments of mixed polarities where motors slide with a velocity $v_m$, a right-(left-)pointing filament moves with a velocity $v_m$ ($-v_m$) relative to any motor. (d)~Crosslinks that remain bound to two antiparallel filaments throughout this dynamics stretch with a velocity $2v_m$ (the top and bottom panels represent the same system with a time interval $\tau$).}
\end{figure*}

We consider a bundle of polar filaments of length $L$ aligned in the $x$-direction and subjected to periodic boundary conditions. The filaments are rigid, ruling out contraction arising from mechanical nonlinearities~\cite{Lenz:2012a}.
A filament may point in the direction of positive or negative $x$, and maintains this polarity throughout the dynamics.
At steady-state, filaments constantly grow from their barbed ends and shrink from their pointed ends at a fixed velocity $v_t$, a phenomenon known as ``treadmilling'' throughout which their length remains constant~\cite{Alberts:2015} [Fig.~\ref{fig:parametrization}(a)]. Motors and crosslinks constantly bind and unbind from filaments, and we denote by $\tau_m$ ($\tau_c$) and $\rho_m^0$ ($\rho_c^0$) the average motor (crosslink) unbinding time and equilibrium density [Fig.~\ref{fig:parametrization}(b)].

Once bound to a filament, motors slide towards its barbed end with a velocity $v_m$. The value of $v_m$ is set by a competition between the propulsive forces of the motors and the restoring forces of the crosslinks, and is to be determined self-consistently at a later stage of the calculation. In a mean-field description (valid for filaments interacting with many neighbors through many motors and crosslinks), this results in the pattern of motion illustrated in Fig.~\ref{fig:parametrization}(c).

Focusing on a single right-pointing filament, the combined effect of motor motion and actin treadmilling implies that motors move with a velocity $v_m-v_t$ relative to the growing barbed end. Denoting by $\xi$ the distance between the motor and the barbed end [Fig.~\ref{fig:parametrization}(c)], this implies that the number of bound motors per unit filament length $\rho_m(\xi,t)$ satisfies the reaction-convection equation:
\begin{equation}\label{eq:motordynamics}
\partial_t\rho_m=-\partial_\xi J_m+\frac{\rho_m^0}{\tau_m}-\frac{\rho_m}{\tau_m},
\end{equation}
where $J_m(\xi,t)=\rho_m(v_t-v_m)$ is the motor current in the reference frame of the barbed end, and ${\rho_m^0}/{\tau_m}$ represents the attachment rate of unbound motors from the surrounding solution. Newly polymerized actin in $\xi=0$ does not yet have any motors bound to it, implying $\rho_m(0,t)=0$ if $v_t>v_m$; likewise $\rho_m(L,t)=0$ if $v_t<v_m$. Motors bound to two filaments of opposing polarities exert forces on each filament, and we denote by $f_m(\xi,t)$ the longitudinal force per unit length exerted by the motors on a right-pointing filament. For independent motors operating close to their stall force (\emph{i.e.}, motors whose velocity is essentially controlled by the external crosslink restoring forces), $f_m(\xi,t)$ is proportional to the local motor density through $f_m(\xi,t)=f_m^0\times[\rho_m(\xi,t)/\rho_m^0]$. Note that motors do not induce internal forces in pairs of filaments with identical polarities, which we thus need not consider here.

The density $\rho_c(\xi,\tau,t)$ of crosslinks of age $\tau$ bound in $\xi$ at time $t$ satisfies the conservation equation
\begin{equation}\label{eq:crosslinkdynamics}
\partial_t\rho_c - \partial_\tau\rho_c=-\partial_\xi J_c+\frac{\rho_c^0\delta(\tau)}{\tau_c}-\frac{\rho_c}{\tau_c},
\end{equation} 
with $\rho_c(0,\tau,t)=\rho_c(\xi,\tau\leq 0,t)=0$. Since the crosslink attachment points do not slide on the actin, their advection relative to the barbed end is entirely due to treadmilling and the crosslink current reads $J_c(\xi,t)=\rho_cv_t$. The term $- \partial_\tau\rho_c$ in Eq.~(\ref{eq:crosslinkdynamics}) can be viewed as an advection term along the coordinate $\tau$, which account for the fact that the age $\tau$ of a bound crosslink increases linearly with time $t$. While attached crosslinks are thus advected towards increasing $\tau$, newly attached crosslinks all have age $\tau=0$ by definition, which we enforce through the delta function in the source term ${\rho_c^0\delta(\tau)}/{\tau_c}$. As motor forces tend to slide filaments of opposing polarities respective to one another, they are opposed by the restoring forces of the crosslinks, which tend to keep filaments stationary with respect to one another. To describe this competition, we assimilate crosslinks to Hookean springs with elastic constant $k_c$. The average extension of a crosslink bound to two antiparallel filaments is equal to zero at the time of its binding (denoted as $\tau=0$), but increases as $2v_m\tau$ as the filaments slide respective to one another [Fig.~\ref{fig:parametrization}(d)]. As each crosslink exerts a Hookean force $-k_c\times(2v_m\tau)$ on the filament, the crosslink force per unit filament length is obtained by summing this force over all filament ages, yielding $f_c(\xi,t)=\int_0^{+\infty}-k_c\times(2v_m\tau)\times\rho_c(\xi,\tau,t)\,\mathrm{d}\tau$.

\begin{figure}[t]
\centering
\includegraphics[width=\columnwidth]{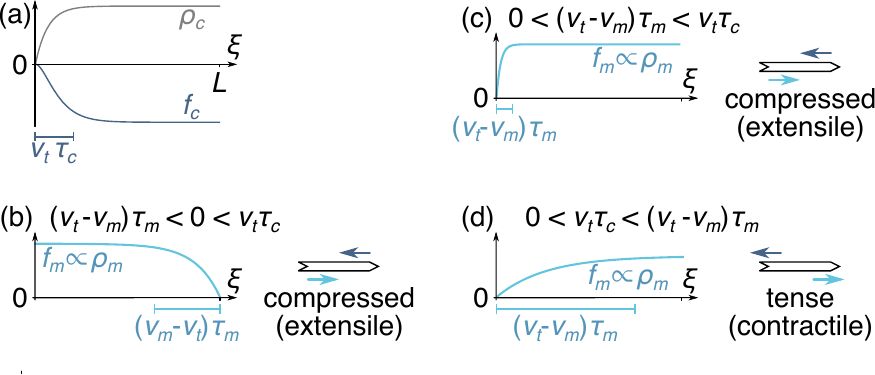}
\caption{\label{fig:profiles}Filament force density profiles as in Eqs.~(\ref{eq:forcedensities}). (a)~The crosslink density $\rho_c$ is suppressed near the barbed end, implying that the crosslink force $f_c<0$ is predominantly applied to the right-hand-side of the filament. (b)~When motors are faster than treadmilling ($v_m>v_t$), they are depleted from the pointed end and right-directed motor forces are predominantly applied on the left-hand side of the filament.
As schematized on the right-hand-side, the fact that the crosslink force (dark blue arrow) is applied more to the right than the motor force (light blue arrow) implies that the filament is under compression. (c)~When $v_t$ barely exceeds $v_m$, motor forces are applied relatively uniformly over the filament, which also results in filament compression. (d)~When $v_t\gg v_m$, the motor depletion zone is larger than the crosslink depletion zone and motor forces are significantly shifted to the right. The filament is tensed in that case.}
\end{figure}

Solving Eqs.~(\ref{eq:motordynamics}-\ref{eq:crosslinkdynamics}), we compute the steady-state force densities exerted by the motors and crosslinks on the filament:
\begin{subequations}\label{eq:forcedensities}
\begin{eqnarray}
f_m(\xi) &=&
\begin{cases} f_m^0\left[1-e^{-{\xi}/{(v_t-v_m)\tau_m}}\right] &\mbox{if } v_t>v_m \\
f_m^0\left[1-e^{-{(L-\xi)}/{(v_m-v_t)\tau_m}}\right] & \mbox{if } v_t<v_m \end{cases}\label{eq:motorforcedensity}\\ 
f_c(\xi) &=&-2k_c\rho_c^0\tau_c{v_m}\left[1-\left(1+\frac{\xi}{v_t\tau_c}\right)e^{-{\xi}/{v_t\tau_c}}\right].\label{eq:crosslinkforcedensity}
\end{eqnarray}
\end{subequations}
Equations~(\ref{eq:forcedensities}) describe a depletion of motors and crosslinks close to the filament ends, with associated depletion lengths $|{v}_t-{v}_m|\tau_m$ and ${v}_t{\tau}_c$, as illustrated in Fig.~\ref{fig:profiles}. The crosslink depletion results from the finite time required to decorate newly polymerized actin with crosslinks, while the motor depletion arises from the time required to dress a newly created filament overlap with motors. This delay may result from the motor binding time as discussed above, or from a rearrangement time required for an already-present motor to properly engage the filament. Provided the filament length is much larger than these depletion lengths, the motor force and crosslink friction asymptotically go to the constant values $f_m^0$ and $-2k_c\rho_c^0\tau_c$ far from the filament ends as the motor and crosslink densities go to their equilibrium values. We denote by $v_m^0=f_m^0/(2k_c\rho_c^0\tau_c)$ the speed at which these asymptotic forces balance each other, which characterizes the hypothetical motion of infinite-length filaments.

\begin{figure}
\includegraphics[width=\columnwidth]{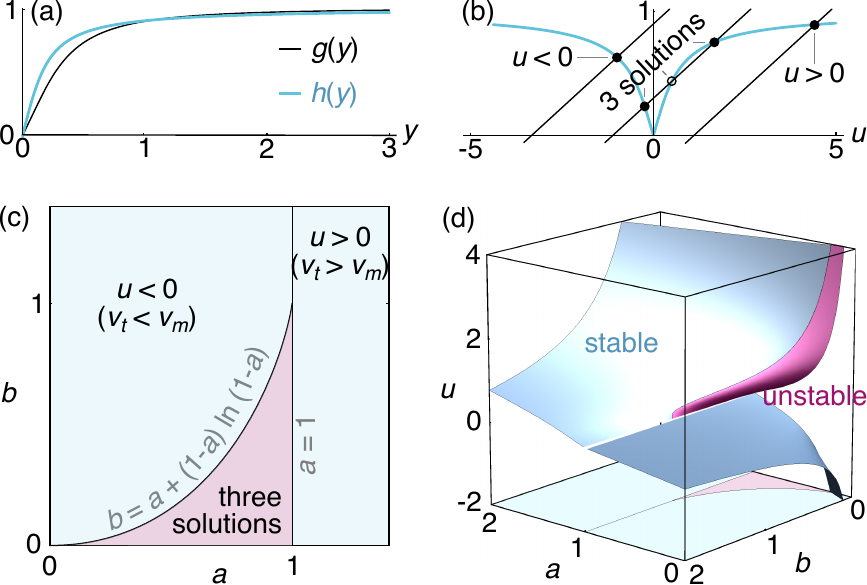}
\caption{\label{fig:velocity}Velocity selection in the bundle. (a)~Profiles of the functions $g(y)$ and $h(y)$, both of which go monotonically from 0 to 1 as $y$ goes from 0 to $+\infty$. (b)~Graphical illustration of the velocity selection condition Eq.~(\ref{eq:forcebalance}) as the intersection between two curves. The blue curve represents the left-hand side of Eq.~(\ref{eq:forcebalance}), and the black lines represent three possible parameter regimes for the right-hand side (here $b=0.27$ and $a=0.1$, $0.7$ and $1.3$ from left to right). Solid (open) circles represent stable (unstable) solutions. (c)~Phase diagram presenting these three regimes as a function of parameters $a$ and $b$. (d)~Values of the rescaled apparent filament velocity $u$ selected by the system, with colors indicating the stability of the solutions. The phase diagram of panel~(c) is reproduced on the bottom face of the plot to facilitate comparisons.}
\end{figure}

By contrast, shorter filaments undergo both a smaller overall driving force and a smaller friction. Depletion thus affects the velocity $v_m$, while $v_m$ itself affects motor depletion as described by Eq.~(\ref{eq:motorforcedensity}). Rescaling all lengths by $v_m^0\tau_m$ and times by $\tau_m$, we henceforth denote dimensionless variants of previously introduced variables with a tilde and determine $\tilde{v}_m$ by demanding that the total force $F=\int_0^L\left[f_m(\xi)+f_c(\xi)\right]\,\text{d}\xi$ exerted on a single filament vanishes. Defining ${u}=(\tilde{v}_t-\tilde{v}_m)/\tilde{L}$, we insert Eqs.~(\ref{eq:forcedensities}) into this condition and obtain a transcendental equation for $u$:
\begin{equation}\label{eq:forcebalance}
|u|(1-e^{-1/|u|})=(1-a)+bu,
\end{equation}
where $a=\tilde{v}_t[1-g(\tilde{v}_t\tilde{\tau}_c/\tilde{L})]$ and $b=\tilde{L}[1-g(\tilde{v}_t\tilde{\tau}_c/\tilde{L})]$ are two constants and $g(y)=2y-(1+2y)e^{-1/y}$ [see Fig.~\ref{fig:velocity}(a)]. As $a>0$ and $b>0$, Eq.~(\ref{eq:forcebalance}) gives rise to three regimes illustrated in Fig.~\ref{fig:velocity}(b-c): one where translocation by the motors is faster than treadmilling ($u<0\Leftrightarrow {v}_m>{v_t}$), one where treadmilling is faster than translocation ($u>0$) and one where one $u<0$ solution coexists with two $u>0$ solutions. We determine the stability of these solutions by perturbing $\tilde{v}_m$ by a small quantity $\delta\tilde{v}_m$ and assessing whether the overall force $F$ exerted on the filament tends to amplify or suppress this perturbation. We find that all unique solutions are stable (\emph{i.e.}, $\partial F/\partial\tilde{v}_m<0$). In the three-solutions regime, the smaller of the two $u>0$ solutions is unstable. The bundle thus chooses one of the other two, resulting in two coexisting stable solutions of opposing signs as illustrated in Fig.~\ref{fig:velocity}(d). As for any first-order (discontinuous) transition, bundles in this parameter regime will select either value of $u$ depending on their initial condition, and any switching from one to the other involves hysteresis.

\begin{figure}[t]
\centering
\includegraphics[width=\columnwidth]{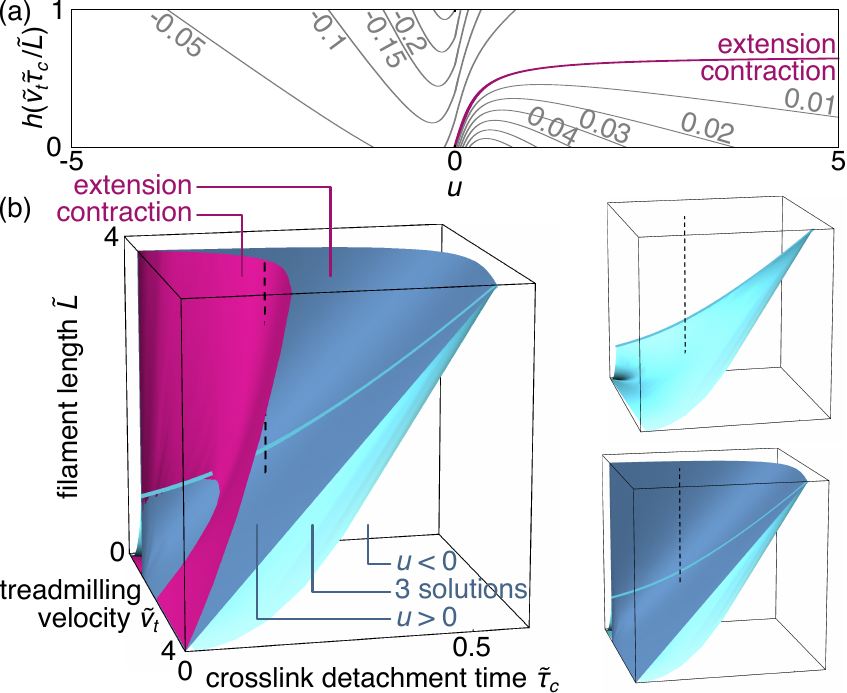}
\caption{\label{fig:tensions}Bundle tension. (a)~Level curves for the dimensionless bundle tension $\tilde{\cal T}$ as a function of the apparent velocity $u$ determined from Eq.~(\ref{eq:forcebalance}) and pictured in Fig.~\ref{fig:velocity}(d), and the ratio $\tilde{v}_t\tilde{\tau}_c/\tilde{L}$. The $\tilde{\cal T}=0$ purple line separates contraction from extension. (b)~Contraction regimes associated with the stable $u>0$ solution (purple surface) and velocity regimes as in Fig.~\ref{fig:velocity}(c) (blue surfaces) as a function of the dimensionless parameters $\tilde{v}_t$, $\tilde{\tau}_c$ and $\tilde{L}$. The blue surfaces are plotted separately on the right to facilitate visualization. As discussed above the ``3 solutions'' regime has coexisting stable $u<0$ and $u>0$ solutions. The light blue line outlines the intersection between the two blue surfaces. The dashed line materializes one set of reasonable experimental parameters (see text), and goes from $u<0$ to $u>0$ through the coexistence (``3 solutions'') region, implying a first-order transition.
By contrast, a similar vertical line shifted to smaller values of $\tilde{v}_t$ would describe a second-order transition.}
\end{figure}

We now turn to the contractile/extensile character of a bundle comprised of $\rho_f$ filaments per unit length. A filament in this bundle is subjected to a total force per unit length $f(\xi)=z[f_m(\xi)+f_c(\xi)]$ at location $\xi$, where $z$ denotes the number of interacting neighbors of a filament. As the filament tension $T(\xi)$ vanishes at the filament ends [$T(0)=T(L)=0$], its tension in $\xi$ thus reads $T(\xi)=\int_0^\xi f(\xi')\,\text{d}\xi'$. The contractile or extensile character of our bundle is revealed by its integrated tension across any $x=\text{constant}$ plane. In thick bundles, this plane is intersected by a large number of filaments (namely $\rho_f L\gg 1$) each intersecting the plane at a random coordinate $\xi$ that is uniformly distributed between $0$ and $L$. As a result, the bundle tension is given by the average ${\cal T}=\rho_f\int_0^L T(\xi){\textrm{d}\xi}$. Defining $\tilde{{\cal T}}={\cal T}/(z\rho_fL^2f_m^0)=\tilde{\cal T}_m+\tilde{\cal T}_c$, the respective contributions of the motors and crosslinks to the dimensionless bundle tension are
\begin{subequations}\label{eq:tensions}
\begin{eqnarray}
\tilde{\cal T}_m&=&
\begin{cases} \frac{1}{2}-u^2+u(1+u)e^{-1/u} &\mbox{if } u>0 \\
\frac{1}{2}-u^2e^{1/u}+u(1+u) & \mbox{if } u<0 \end{cases}\\
\tilde{\cal T}_c&=&\frac{|u|\left(1-e^{-1/|u|}\right)-1}{4}\left[2+h\left(\frac{\tilde{v}_t\tilde{\tau}_c}{\tilde{L}}\right)\right],
\end{eqnarray}
\end{subequations}
where the function $h(y)=[4y-12y^2+(2+8y+12y^2)e^{-1/y}]/[(1-2y)+(1+2y)e^{-1/y}]$ is illustrated in Fig.~\ref{fig:velocity}(a). As shown in Fig.~\ref{fig:tensions}(a), these expressions can result in either sign for ${\cal T}$ depending on the values of $u$ and $h({\tilde{v}_t\tilde{\tau}_c}/{\tilde{L}})$. As the periodic boundary conditions used here confine the bundle to a fixed length, a bundle with a propensity to extend develops a negative tension ${\cal T}<0$ (\emph{i.e.}, is compressed), while ${\cal T}>0$ denotes a contractile (tense) bundle. These two behaviors respectively correspond to the situations illustrated in Fig.~\ref{fig:profiles}(b-c) and Fig.~\ref{fig:profiles}(d). We illustrate the regimes in Fig.~\ref{fig:tensions}(b) as a function of the original dimensionless parameters $\tilde{v}_t$, $\tilde{\tau}_c$ and $\tilde{L}$. As some parameter values yield coexisting metastable values of $u$, so can they allow for both contractile and extensile steady states. However, despite this ambiguity at intermediate parameter values, Fig.~\ref{fig:tensions}(b) shows that the self-organization mechanism investigated here results in unambiguous extension for broad ranges of parameters.

This transition from contractile to extensile behavior upon an increase in the crosslink detachment time $\tau_c$ can be rationalized by an enlarged crosslink depletion zone in the vicinity of the filament barbed ends (Fig.~\ref{fig:profiles}). This implies a localization of crosslinks towards the filament pointed ends, resulting in an extensile ``anti-sarcomere'' organization [Fig.~\ref{fig:mechanism}(c), right], in contrast with the contractile ``sarcomere'' structures [Fig.~\ref{fig:mechanism}(c), left] found in our highly organized striated muscle. The emphasis of this mechanism on barbed end depletion suggests that it will not be significantly affected if pointed end assembly proceeds through severing~\cite{Theriot:1997} rather than depolymerization. We also predict another transition, whereby a further increase of $\tau_c$ in the extensile phase causes the variable $u\propto v_t-v_m$ to change sign through a first-order (for small $\tilde{L}$) or a second-order (for large $\tilde{L}$) transition [Fig.~\ref{fig:tensions}(b)]. Indeed, the enhanced crosslink depletion associated with a large $\tau_c$ tends to reduce the friction between filaments, resulting in faster motor motion and thus in a situation where motor sliding outpaces treadmilling ($u<0$). Both transitions could be directly observed by manipulating the actin dynamics or motor composition in current \emph{in vitro} assays~\cite{Thoresen:2013,Reymann:2012,Murrell:2012} and possibly in cells~\cite{Mendes:2012}. Such changes could also be at work in smooth muscle, where the number of myosins in individual thick filaments is regulated dynamically~\cite{Seow:2005}. The experimental relevance of these transitions is illustrated by a dashed line in Fig.~\ref{fig:tensions}(b), which shows that both transitions can be probed by varying $L$ between $250\,\text{nm}$ and $1\,\mu\text{m}$ while holding $v_m^0=50\,\text{nm}\cdot\text{s}^{-1}$, $\tau_m=5\,\text{s}$~\cite{Erdmann:2013}, $v_t=100\,\text{nm}\cdot\text{s}^{-1}$~\cite{Howard:2001aa}, and $\tau_c=1\,\text{s}$~\cite{Miyata:1996} fixed. In addition, the magnitude of the forces and velocities predicted here are on par with those found \emph{in vivo}, \emph{e.g.}, in the cytokinetic ring of fission yeast. Indeed, setting $L=1.4\,\mu\text{m}$, $\rho_fL=20$, $f_m^0\simeq 7.2\times 10^{-6}\,\text{N}\cdot\text{m}^{-1}$~\cite{Wu:2005}, $k_c\simeq 3\times 10^{-4}\,\text{N}\cdot\text{m}^{-1}$~\cite{Rief:1999}, $z=6$ as in a hexagonal packing and $\tilde{\cal T}\simeq 0.1$, we find a contractile force ${\cal T}\simeq 120\,\text{pN}$ comparable with the force required for fission ${\cal T}_\text{req}\simeq 160\,\text{pN}$ extrapolated from the required cleavage force in echinoderm eggs~\cite{Rappaport:1967} to a yeast ring with radius $1\,\mu\text{m}$~\cite{Zumdieck:2007}. We also find a characteristic velocity $v_m^0\simeq 5\,\text{nm}\cdot\text{s}^{-1}$ similar to that of ring contraction ($\simeq 3\text{-}4\,\text{nm}\cdot\text{s}^{-1}$).

Overall, our prediction that self-organized force generation entails a robust extensile regime provides a stringent test to validate or invalidate this model in specific experiments. For instance, the bundles of Refs.~\cite{Thoresen:2011,Thoresen:2013} contract despite the fact that $v_t=0$, which contradicts the self-organization prediction and thus validates the fact that they are dominated by mechanical nonlinearities. Conversely, stiff microtubules systems where buckling nonlinearities are strongly suppressed extend \emph{in vitro} when the filament polymerization/depolymerization dynamics is blocked~\cite{Sanchez:2012,Keber:2014} and contract in more complex \emph{in vivo} situations~\cite{Foster:2015}, consistent with the self-organization model. In addition, both extension and contraction have also been reported in stiff, non-buckling actin bundles~\cite{Stam:2017}. Beyond pure contraction or extension, transitions between these two possibly coexisting (as in the multiple-solution regime of Fig.~\ref{fig:velocity}) metastable states could help understand several \emph{in vivo} behaviors involving alternating contractions and expansions of the actomyosin cortex. This includes cell area oscillations observed during \emph{Drosophila}, \emph{C. elegans}, and \emph{Xenopus} development~\cite{Roh-Johnson:2012,Kim:2011a} or propagating actomyosin contractility waves~\cite{Allard:2013}. It would also be interesting to see how the mechanisms described here apply to the more complex geometry of two- or three-dimensional actomyosin assemblies~\cite{Lenz:2014}, and to connect our self-organization mechanisms to the onset of positional ordering in muscle-like bundles~\cite{Friedrich:2012,Hu:2017}. 
Finally, the fundamental principles for the dynamical depletion of motors and crosslinks described here could serve as guiding principles in our nascent understanding of self-organized contractility in the cytoskeleton.

\begin{acknowledgments}
I thank Alex Mogilner for sharing Ref.~\cite{Oelz:2015} before publication, Pierre Ronceray for enlightening discussions and Samuel Cazayus-Claverie, Michael Murrell, Guglielmo Saggiorato and Danny Seara for comments on the manuscript. This work was supported by Marie Curie Integration Grant PCIG12-GA-2012-334053, ``Investissements d'Avenir'' LabEx PALM (ANR-10-LABX-0039-PALM), ANR grant ANR-15-CE13-0004-03 and ERC Starting Grant 677532. My group belongs to the CNRS consortium CellTiss.
\end{acknowledgments}


\begin{thebibliography}{40}
\expandafter\ifx\csname natexlab\endcsname\relax\def\natexlab#1{#1}\fi
\expandafter\ifx\csname bibnamefont\endcsname\relax
  \def\bibnamefont#1{#1}\fi
\expandafter\ifx\csname bibfnamefont\endcsname\relax
  \def\bibfnamefont#1{#1}\fi
\expandafter\ifx\csname citenamefont\endcsname\relax
  \def\citenamefont#1{#1}\fi
\expandafter\ifx\csname url\endcsname\relax
  \def\url#1{\texttt{#1}}\fi
\expandafter\ifx\csname urlprefix\endcsname\relax\def\urlprefix{URL }\fi
\providecommand{\bibinfo}[2]{#2}
\providecommand{\eprint}[2][]{\url{#2}}

\bibitem[{\citenamefont{Verkhovsky et~al.}(1995)\citenamefont{Verkhovsky,
  Svitkina, and Borisy}}]{Verkhovsky:1995}
\bibinfo{author}{\bibfnamefont{A.~B.} \bibnamefont{Verkhovsky}},
  \bibinfo{author}{\bibfnamefont{T.~M.} \bibnamefont{Svitkina}},
  \bibnamefont{and} \bibinfo{author}{\bibfnamefont{G.~G.}
  \bibnamefont{Borisy}}, \bibinfo{journal}{J. Cell Biol.}
  \textbf{\bibinfo{volume}{131}}, \bibinfo{pages}{989} (\bibinfo{year}{1995}).

\bibitem[{\citenamefont{Cramer et~al.}(1997)\citenamefont{Cramer, Siebert, and
  Mitchison}}]{Cramer:1997}
\bibinfo{author}{\bibfnamefont{L.~P.} \bibnamefont{Cramer}},
  \bibinfo{author}{\bibfnamefont{M.}~\bibnamefont{Siebert}}, \bibnamefont{and}
  \bibinfo{author}{\bibfnamefont{T.~J.} \bibnamefont{Mitchison}},
  \bibinfo{journal}{J. Cell Biol.} \textbf{\bibinfo{volume}{136}},
  \bibinfo{pages}{1287} (\bibinfo{year}{1997}).

\bibitem[{\citenamefont{Medalia et~al.}(2002)\citenamefont{Medalia, Weber,
  Frangakis, Nicastro, Gerisch, and Baumeister}}]{Medalia:2002}
\bibinfo{author}{\bibfnamefont{O.}~\bibnamefont{Medalia}},
  \bibinfo{author}{\bibfnamefont{I.}~\bibnamefont{Weber}},
  \bibinfo{author}{\bibfnamefont{A.~S.} \bibnamefont{Frangakis}},
  \bibinfo{author}{\bibfnamefont{D.}~\bibnamefont{Nicastro}},
  \bibinfo{author}{\bibfnamefont{G.}~\bibnamefont{Gerisch}}, \bibnamefont{and}
  \bibinfo{author}{\bibfnamefont{W.}~\bibnamefont{Baumeister}},
  \bibinfo{journal}{Science} \textbf{\bibinfo{volume}{298}},
  \bibinfo{pages}{1209} (\bibinfo{year}{2002}).

\bibitem[{\citenamefont{Kamasaki et~al.}(2007)\citenamefont{Kamasaki, Osumi,
  and Mabuchi}}]{Kamasaki:2007}
\bibinfo{author}{\bibfnamefont{T.}~\bibnamefont{Kamasaki}},
  \bibinfo{author}{\bibfnamefont{M.}~\bibnamefont{Osumi}}, \bibnamefont{and}
  \bibinfo{author}{\bibfnamefont{I.}~\bibnamefont{Mabuchi}},
  \bibinfo{journal}{J. Cell Biol.} \textbf{\bibinfo{volume}{178}},
  \bibinfo{pages}{765} (\bibinfo{year}{2007}).

\bibitem[{\citenamefont{Murrell et~al.}(2015)\citenamefont{Murrell, Oakes,
  Lenz, and Gardel}}]{Murrell:2015}
\bibinfo{author}{\bibfnamefont{M.}~\bibnamefont{Murrell}},
  \bibinfo{author}{\bibfnamefont{P.~W.} \bibnamefont{Oakes}},
  \bibinfo{author}{\bibfnamefont{M.}~\bibnamefont{Lenz}}, \bibnamefont{and}
  \bibinfo{author}{\bibfnamefont{M.~L.} \bibnamefont{Gardel}},
  \bibinfo{journal}{Nat. Rev. Mol. Cell Biol.} \textbf{\bibinfo{volume}{16}},
  \bibinfo{pages}{486} (\bibinfo{year}{2015}).

\bibitem[{\citenamefont{Hatano}(1994)}]{Hatano:1994}
\bibinfo{author}{\bibfnamefont{S.}~\bibnamefont{Hatano}},
  \bibinfo{journal}{Int. Rev. Cytology} \textbf{\bibinfo{volume}{156}},
  \bibinfo{pages}{199} (\bibinfo{year}{1994}).

\bibitem[{\citenamefont{Sekimoto and Nakazawa}(1998)}]{Sekimoto:1998}
\bibinfo{author}{\bibfnamefont{K.}~\bibnamefont{Sekimoto}} \bibnamefont{and}
  \bibinfo{author}{\bibfnamefont{H.}~\bibnamefont{Nakazawa}},
  \emph{\bibinfo{title}{Current Topics in Physics}} (\bibinfo{publisher}{World
  Scientific}, \bibinfo{address}{Singapore}, \bibinfo{year}{1998}),
  vol.~\bibinfo{volume}{1}, pp. \bibinfo{pages}{394--405}.

\bibitem[{\citenamefont{Lenz et~al.}(2012{\natexlab{a}})\citenamefont{Lenz,
  Gardel, and Dinner}}]{Lenz:2012a}
\bibinfo{author}{\bibfnamefont{M.}~\bibnamefont{Lenz}},
  \bibinfo{author}{\bibfnamefont{M.~L.} \bibnamefont{Gardel}},
  \bibnamefont{and} \bibinfo{author}{\bibfnamefont{A.~R.}
  \bibnamefont{Dinner}}, \bibinfo{journal}{New J. Phys.}
  \textbf{\bibinfo{volume}{14}}, \bibinfo{pages}{033037}
  (\bibinfo{year}{2012}{\natexlab{a}}).

\bibitem[{\citenamefont{Pinto et~al.}(2013)\citenamefont{Pinto, Rubinstein, and
  Li}}]{Pinto:2013}
\bibinfo{author}{\bibfnamefont{I.~M.} \bibnamefont{Pinto}},
  \bibinfo{author}{\bibfnamefont{B.}~\bibnamefont{Rubinstein}},
  \bibnamefont{and} \bibinfo{author}{\bibfnamefont{R.}~\bibnamefont{Li}},
  \bibinfo{journal}{Biophys. J.} \textbf{\bibinfo{volume}{105}},
  \bibinfo{pages}{547} (\bibinfo{year}{2013}).

\bibitem[{\citenamefont{Dasanayake et~al.}(2011)\citenamefont{Dasanayake,
  Michalski, and Carlsson}}]{Dasanayake:2011}
\bibinfo{author}{\bibfnamefont{N.~L.} \bibnamefont{Dasanayake}},
  \bibinfo{author}{\bibfnamefont{P.~J.} \bibnamefont{Michalski}},
  \bibnamefont{and} \bibinfo{author}{\bibfnamefont{A.~E.}
  \bibnamefont{Carlsson}}, \bibinfo{journal}{Phys. Rev. Lett.}
  \textbf{\bibinfo{volume}{107}}, \bibinfo{pages}{118101}
  (\bibinfo{year}{2011}).

\bibitem[{\citenamefont{Lenz et~al.}(2012{\natexlab{b}})\citenamefont{Lenz,
  Thoresen, Gardel, and Dinner}}]{Lenz:2012}
\bibinfo{author}{\bibfnamefont{M.}~\bibnamefont{Lenz}},
  \bibinfo{author}{\bibfnamefont{T.}~\bibnamefont{Thoresen}},
  \bibinfo{author}{\bibfnamefont{M.~L.} \bibnamefont{Gardel}},
  \bibnamefont{and} \bibinfo{author}{\bibfnamefont{A.~R.}
  \bibnamefont{Dinner}}, \bibinfo{journal}{Phys. Rev. Lett.}
  \textbf{\bibinfo{volume}{108}}, \bibinfo{pages}{238107}
  (\bibinfo{year}{2012}{\natexlab{b}}).

\bibitem[{\citenamefont{Ronceray et~al.}(2016)\citenamefont{Ronceray,
  Broedersz, and Lenz}}]{Ronceray:2016}
\bibinfo{author}{\bibfnamefont{P.}~\bibnamefont{Ronceray}},
  \bibinfo{author}{\bibfnamefont{C.}~\bibnamefont{Broedersz}},
  \bibnamefont{and} \bibinfo{author}{\bibfnamefont{M.}~\bibnamefont{Lenz}},
  \bibinfo{journal}{Proc. Natl. Acad. Sci. U.S.A.}
  \textbf{\bibinfo{volume}{113}}, \bibinfo{pages}{2827} (\bibinfo{year}{2016}).

\bibitem[{\citenamefont{Kruse and Sekimoto}(2002)}]{Kruse:2002}
\bibinfo{author}{\bibfnamefont{K.}~\bibnamefont{Kruse}} \bibnamefont{and}
  \bibinfo{author}{\bibfnamefont{K.}~\bibnamefont{Sekimoto}},
  \bibinfo{journal}{Phys. Rev. E} \textbf{\bibinfo{volume}{66}},
  \bibinfo{pages}{031904} (\bibinfo{year}{2002}).

\bibitem[{\citenamefont{Zumdieck et~al.}(2007)\citenamefont{Zumdieck, Kruse,
  Bringmann, Hyman, and J\"ulicher}}]{Zumdieck:2007}
\bibinfo{author}{\bibfnamefont{A.}~\bibnamefont{Zumdieck}},
  \bibinfo{author}{\bibfnamefont{K.}~\bibnamefont{Kruse}},
  \bibinfo{author}{\bibfnamefont{H.}~\bibnamefont{Bringmann}},
  \bibinfo{author}{\bibfnamefont{A.~A.} \bibnamefont{Hyman}}, \bibnamefont{and}
  \bibinfo{author}{\bibfnamefont{F.}~\bibnamefont{J\"ulicher}},
  \bibinfo{journal}{{PL}o{S} One} \textbf{\bibinfo{volume}{2}},
  \bibinfo{pages}{e696} (\bibinfo{year}{2007}).

\bibitem[{\citenamefont{Zemel and Mogilner}(2009)}]{Zemel:2009}
\bibinfo{author}{\bibfnamefont{A.}~\bibnamefont{Zemel}} \bibnamefont{and}
  \bibinfo{author}{\bibfnamefont{A.}~\bibnamefont{Mogilner}},
  \bibinfo{journal}{Phys. Chem. Chem. Phys.} \textbf{\bibinfo{volume}{11}},
  \bibinfo{pages}{4821} (\bibinfo{year}{2009}).

\bibitem[{\citenamefont{Oelz et~al.}(2015)\citenamefont{Oelz, Rubinstein, and
  Mogilner}}]{Oelz:2015}
\bibinfo{author}{\bibfnamefont{D.~B.} \bibnamefont{Oelz}},
  \bibinfo{author}{\bibfnamefont{B.~Y.} \bibnamefont{Rubinstein}},
  \bibnamefont{and} \bibinfo{author}{\bibfnamefont{A.}~\bibnamefont{Mogilner}},
  \bibinfo{journal}{Biophys. J.} \textbf{\bibinfo{volume}{109}},
  \bibinfo{pages}{1818} (\bibinfo{year}{2015}).

\bibitem[{\citenamefont{Alberts et~al.}(2015)\citenamefont{Alberts, Johnson,
  Lewis, Morgan, Raff, Roberts, and Walter}}]{Alberts:2015}
\bibinfo{author}{\bibfnamefont{B.}~\bibnamefont{Alberts}},
  \bibinfo{author}{\bibfnamefont{A.}~\bibnamefont{Johnson}},
  \bibinfo{author}{\bibfnamefont{J.}~\bibnamefont{Lewis}},
  \bibinfo{author}{\bibfnamefont{D.}~\bibnamefont{Morgan}},
  \bibinfo{author}{\bibfnamefont{M.}~\bibnamefont{Raff}},
  \bibinfo{author}{\bibfnamefont{K.}~\bibnamefont{Roberts}}, \bibnamefont{and}
  \bibinfo{author}{\bibfnamefont{P.}~\bibnamefont{Walter}},
  \emph{\bibinfo{title}{Molecular biology of the cell}}
  (\bibinfo{publisher}{Garland Science}, \bibinfo{year}{2015}),
  \bibinfo{edition}{6th} ed.

\bibitem[{\citenamefont{Theriot}(1997)}]{Theriot:1997}
\bibinfo{author}{\bibfnamefont{J.~A.} \bibnamefont{Theriot}},
  \bibinfo{journal}{J Cell Biol} \textbf{\bibinfo{volume}{136}},
  \bibinfo{pages}{1165} (\bibinfo{year}{1997}).

\bibitem[{\citenamefont{Thoresen et~al.}(2013)\citenamefont{Thoresen, Lenz, and
  Gardel}}]{Thoresen:2013}
\bibinfo{author}{\bibfnamefont{T.}~\bibnamefont{Thoresen}},
  \bibinfo{author}{\bibfnamefont{M.}~\bibnamefont{Lenz}}, \bibnamefont{and}
  \bibinfo{author}{\bibfnamefont{M.~L.} \bibnamefont{Gardel}},
  \bibinfo{journal}{Biophys. J.} \textbf{\bibinfo{volume}{104}},
  \bibinfo{pages}{655} (\bibinfo{year}{2013}).

\bibitem[{\citenamefont{Reymann et~al.}(2012)\citenamefont{Reymann,
  Boujemaa-Paterski, Martiel, Gu\'erin, Cao, Chin, Cruz, Th\'ery, and
  Blanchoin}}]{Reymann:2012}
\bibinfo{author}{\bibfnamefont{A.-C.} \bibnamefont{Reymann}},
  \bibinfo{author}{\bibfnamefont{R.}~\bibnamefont{Boujemaa-Paterski}},
  \bibinfo{author}{\bibfnamefont{J.-L.} \bibnamefont{Martiel}},
  \bibinfo{author}{\bibfnamefont{C.}~\bibnamefont{Gu\'erin}},
  \bibinfo{author}{\bibfnamefont{W.}~\bibnamefont{Cao}},
  \bibinfo{author}{\bibfnamefont{H.~F.} \bibnamefont{Chin}},
  \bibinfo{author}{\bibfnamefont{E.~M. D.~L.} \bibnamefont{Cruz}},
  \bibinfo{author}{\bibfnamefont{M.}~\bibnamefont{Th\'ery}}, \bibnamefont{and}
  \bibinfo{author}{\bibfnamefont{L.}~\bibnamefont{Blanchoin}},
  \bibinfo{journal}{Science} \textbf{\bibinfo{volume}{336}},
  \bibinfo{pages}{1310} (\bibinfo{year}{2012}).

\bibitem[{\citenamefont{Murrell and Gardel}(2012)}]{Murrell:2012}
\bibinfo{author}{\bibfnamefont{M.}~\bibnamefont{Murrell}} \bibnamefont{and}
  \bibinfo{author}{\bibfnamefont{M.~L.} \bibnamefont{Gardel}},
  \bibinfo{journal}{Proc. Natl. Acad. Sci. U.S.A.}
  \textbf{\bibinfo{volume}{109}}, \bibinfo{pages}{20820}
  (\bibinfo{year}{2012}).

\bibitem[{\citenamefont{Pinto et~al.}(2012)\citenamefont{Pinto, Rubinstein,
  Kucharavy, Unruh, and Li}}]{Mendes:2012}
\bibinfo{author}{\bibfnamefont{I.~M.} \bibnamefont{Pinto}},
  \bibinfo{author}{\bibfnamefont{B.}~\bibnamefont{Rubinstein}},
  \bibinfo{author}{\bibfnamefont{A.}~\bibnamefont{Kucharavy}},
  \bibinfo{author}{\bibfnamefont{J.~R.} \bibnamefont{Unruh}}, \bibnamefont{and}
  \bibinfo{author}{\bibfnamefont{R.}~\bibnamefont{Li}}, \bibinfo{journal}{Dev.
  Cell.} \textbf{\bibinfo{volume}{22}}, \bibinfo{pages}{1247}
  (\bibinfo{year}{2012}).

\bibitem[{\citenamefont{Seow}(2005)}]{Seow:2005}
\bibinfo{author}{\bibfnamefont{C.~Y.} \bibnamefont{Seow}},
  \bibinfo{journal}{Am. J. Physiol.-Cell Physiol.}
  \textbf{\bibinfo{volume}{289}}, \bibinfo{pages}{C1363}
  (\bibinfo{year}{2005}).

\bibitem[{\citenamefont{Erdmann et~al.}(2013)\citenamefont{Erdmann, Albert, and
  Schwarz}}]{Erdmann:2013}
\bibinfo{author}{\bibfnamefont{T.}~\bibnamefont{Erdmann}},
  \bibinfo{author}{\bibfnamefont{P.~J.} \bibnamefont{Albert}},
  \bibnamefont{and} \bibinfo{author}{\bibfnamefont{U.~S.}
  \bibnamefont{Schwarz}}, \bibinfo{journal}{J. Chem. Phys.}
  \textbf{\bibinfo{volume}{139}}, \bibinfo{pages}{175104}
  (\bibinfo{year}{2013}).

\bibitem[{\citenamefont{Howard}(2001)}]{Howard:2001aa}
\bibinfo{author}{\bibfnamefont{J.}~\bibnamefont{Howard}},
  \emph{\bibinfo{title}{Mechanics of Motor Proteins and the Cytoskeleton}}
  (\bibinfo{publisher}{Sinauer Associates}, \bibinfo{address}{Sunderland, MA},
  \bibinfo{year}{2001}).

\bibitem[{\citenamefont{Miyata et~al.}(1996)\citenamefont{Miyata, Yasuda, and
  Kinosita}}]{Miyata:1996}
\bibinfo{author}{\bibfnamefont{H.}~\bibnamefont{Miyata}},
  \bibinfo{author}{\bibfnamefont{R.}~\bibnamefont{Yasuda}}, \bibnamefont{and}
  \bibinfo{author}{\bibfnamefont{K.~J.} \bibnamefont{Kinosita}},
  \bibinfo{journal}{Biochim. Biophys. Acta} \textbf{\bibinfo{volume}{1290}},
  \bibinfo{pages}{83} (\bibinfo{year}{1996}).

\bibitem[{\citenamefont{Wu and Pollard}(2005)}]{Wu:2005}
\bibinfo{author}{\bibfnamefont{J.-Q.} \bibnamefont{Wu}} \bibnamefont{and}
  \bibinfo{author}{\bibfnamefont{T.~D.} \bibnamefont{Pollard}},
  \bibinfo{journal}{Science} \textbf{\bibinfo{volume}{310}},
  \bibinfo{pages}{310} (\bibinfo{year}{2005}).

\bibitem[{\citenamefont{Rief et~al.}(1999)\citenamefont{Rief, Pascual, Saraste,
  and Gaub}}]{Rief:1999}
\bibinfo{author}{\bibfnamefont{M.}~\bibnamefont{Rief}},
  \bibinfo{author}{\bibfnamefont{J.}~\bibnamefont{Pascual}},
  \bibinfo{author}{\bibfnamefont{M.}~\bibnamefont{Saraste}}, \bibnamefont{and}
  \bibinfo{author}{\bibfnamefont{H.~E.} \bibnamefont{Gaub}},
  \bibinfo{journal}{J. Mol. Biol.} \textbf{\bibinfo{volume}{286}},
  \bibinfo{pages}{553} (\bibinfo{year}{1999}).

\bibitem[{\citenamefont{Rappaport}(1967)}]{Rappaport:1967}
\bibinfo{author}{\bibfnamefont{R.}~\bibnamefont{Rappaport}},
  \bibinfo{journal}{Science} \textbf{\bibinfo{volume}{156}},
  \bibinfo{pages}{1241} (\bibinfo{year}{1967}).

\bibitem[{\citenamefont{Thoresen et~al.}(2011)\citenamefont{Thoresen, Lenz, and
  Gardel}}]{Thoresen:2011}
\bibinfo{author}{\bibfnamefont{T.}~\bibnamefont{Thoresen}},
  \bibinfo{author}{\bibfnamefont{M.}~\bibnamefont{Lenz}}, \bibnamefont{and}
  \bibinfo{author}{\bibfnamefont{M.~L.} \bibnamefont{Gardel}},
  \bibinfo{journal}{Biophys. J.} \textbf{\bibinfo{volume}{100}},
  \bibinfo{pages}{2698} (\bibinfo{year}{2011}).

\bibitem[{\citenamefont{Sanchez et~al.}(2012)\citenamefont{Sanchez, Chen,
  DeCamp, Heymann, and Dogic}}]{Sanchez:2012}
\bibinfo{author}{\bibfnamefont{T.}~\bibnamefont{Sanchez}},
  \bibinfo{author}{\bibfnamefont{D.~T.~N.} \bibnamefont{Chen}},
  \bibinfo{author}{\bibfnamefont{S.~J.} \bibnamefont{DeCamp}},
  \bibinfo{author}{\bibfnamefont{M.}~\bibnamefont{Heymann}}, \bibnamefont{and}
  \bibinfo{author}{\bibfnamefont{Z.}~\bibnamefont{Dogic}},
  \bibinfo{journal}{Nature} \textbf{\bibinfo{volume}{491}},
  \bibinfo{pages}{431} (\bibinfo{year}{2012}).

\bibitem[{\citenamefont{Keber et~al.}(2014)\citenamefont{Keber, Loiseau,
  Sanchez, DeCamp, Giomi, Bowick, Marchetti, Dogic, and Bausch}}]{Keber:2014}
\bibinfo{author}{\bibfnamefont{F.~C.} \bibnamefont{Keber}},
  \bibinfo{author}{\bibfnamefont{E.}~\bibnamefont{Loiseau}},
  \bibinfo{author}{\bibfnamefont{T.}~\bibnamefont{Sanchez}},
  \bibinfo{author}{\bibfnamefont{S.~J.} \bibnamefont{DeCamp}},
  \bibinfo{author}{\bibfnamefont{L.}~\bibnamefont{Giomi}},
  \bibinfo{author}{\bibfnamefont{M.~J.} \bibnamefont{Bowick}},
  \bibinfo{author}{\bibfnamefont{M.~C.} \bibnamefont{Marchetti}},
  \bibinfo{author}{\bibfnamefont{Z.}~\bibnamefont{Dogic}}, \bibnamefont{and}
  \bibinfo{author}{\bibfnamefont{A.~R.} \bibnamefont{Bausch}},
  \bibinfo{journal}{Science} \textbf{\bibinfo{volume}{345}},
  \bibinfo{pages}{1135} (\bibinfo{year}{2014}).

\bibitem[{\citenamefont{Foster et~al.}(2015)\citenamefont{Foster, F\''urthauer,
  Shelley, and Needleman}}]{Foster:2015}
\bibinfo{author}{\bibfnamefont{P.~J.} \bibnamefont{Foster}},
  \bibinfo{author}{\bibfnamefont{S.}~\bibnamefont{F\''urthauer}},
  \bibinfo{author}{\bibfnamefont{M.~J.} \bibnamefont{Shelley}},
  \bibnamefont{and} \bibinfo{author}{\bibfnamefont{D.~J.}
  \bibnamefont{Needleman}}, \bibinfo{journal}{{eLife}}
  \textbf{\bibinfo{volume}{4}}, \bibinfo{pages}{e10837} (\bibinfo{year}{2015}).

\bibitem[{\citenamefont{Stam et~al.}(2017)\citenamefont{Stam, Freedman,
  Banerjee, Weirich, Dinner, and Gardel}}]{Stam:2017}
\bibinfo{author}{\bibfnamefont{S.}~\bibnamefont{Stam}},
  \bibinfo{author}{\bibfnamefont{S.~L.} \bibnamefont{Freedman}},
  \bibinfo{author}{\bibfnamefont{S.}~\bibnamefont{Banerjee}},
  \bibinfo{author}{\bibfnamefont{K.~L.} \bibnamefont{Weirich}},
  \bibinfo{author}{\bibfnamefont{A.~R.} \bibnamefont{Dinner}},
  \bibnamefont{and} \bibinfo{author}{\bibfnamefont{M.~L.}
  \bibnamefont{Gardel}}, \bibinfo{journal}{bio{RX}iv} p.
  \bibinfo{pages}{141796} (\bibinfo{year}{2017}).

\bibitem[{\citenamefont{Roh-Johnson et~al.}(2012)\citenamefont{Roh-Johnson,
  Shemer, Higgins, McClellan, Werts, Tulu, Gao, Betzig, Kiehart, and
  Goldstein}}]{Roh-Johnson:2012}
\bibinfo{author}{\bibfnamefont{M.}~\bibnamefont{Roh-Johnson}},
  \bibinfo{author}{\bibfnamefont{G.}~\bibnamefont{Shemer}},
  \bibinfo{author}{\bibfnamefont{C.~D.} \bibnamefont{Higgins}},
  \bibinfo{author}{\bibfnamefont{J.~H.} \bibnamefont{McClellan}},
  \bibinfo{author}{\bibfnamefont{A.~D.} \bibnamefont{Werts}},
  \bibinfo{author}{\bibfnamefont{U.~S.} \bibnamefont{Tulu}},
  \bibinfo{author}{\bibfnamefont{L.}~\bibnamefont{Gao}},
  \bibinfo{author}{\bibfnamefont{E.}~\bibnamefont{Betzig}},
  \bibinfo{author}{\bibfnamefont{D.~P.} \bibnamefont{Kiehart}},
  \bibnamefont{and}
  \bibinfo{author}{\bibfnamefont{B.}~\bibnamefont{Goldstein}},
  \bibinfo{journal}{Science} \textbf{\bibinfo{volume}{335}},
  \bibinfo{pages}{1232} (\bibinfo{year}{2012}).

\bibitem[{\citenamefont{Kim and Davidson}(2011)}]{Kim:2011a}
\bibinfo{author}{\bibfnamefont{H.~Y.} \bibnamefont{Kim}} \bibnamefont{and}
  \bibinfo{author}{\bibfnamefont{L.~A.} \bibnamefont{Davidson}},
  \bibinfo{journal}{J Cell Sci} \textbf{\bibinfo{volume}{124}},
  \bibinfo{pages}{635} (\bibinfo{year}{2011}).

\bibitem[{\citenamefont{Allard and Mogilner}(2013)}]{Allard:2013}
\bibinfo{author}{\bibfnamefont{J.}~\bibnamefont{Allard}} \bibnamefont{and}
  \bibinfo{author}{\bibfnamefont{A.}~\bibnamefont{Mogilner}},
  \bibinfo{journal}{Curr. Opin. Cell Biol.} \textbf{\bibinfo{volume}{25}},
  \bibinfo{pages}{107} (\bibinfo{year}{2013}).

\bibitem[{\citenamefont{Lenz}(2014)}]{Lenz:2014}
\bibinfo{author}{\bibfnamefont{M.}~\bibnamefont{Lenz}}, \bibinfo{journal}{Phys.
  Rev. X} \textbf{\bibinfo{volume}{4}}, \bibinfo{pages}{041002}
  (\bibinfo{year}{2014}).

\bibitem[{\citenamefont{Friedrich et~al.}(2012)\citenamefont{Friedrich,
  Fischer-Friedrich, Gov, and Safran}}]{Friedrich:2012}
\bibinfo{author}{\bibfnamefont{B.~M.} \bibnamefont{Friedrich}},
  \bibinfo{author}{\bibfnamefont{E.}~\bibnamefont{Fischer-Friedrich}},
  \bibinfo{author}{\bibfnamefont{N.~S.} \bibnamefont{Gov}}, \bibnamefont{and}
  \bibinfo{author}{\bibfnamefont{S.~A.} \bibnamefont{Safran}},
  \bibinfo{journal}{{PL}o{S} Comput. Biol.} \textbf{\bibinfo{volume}{8}},
  \bibinfo{pages}{e1002544} (\bibinfo{year}{2012}).

\bibitem[{\citenamefont{Hu et~al.}(2017)\citenamefont{Hu, Dasbiswas, Guo, Tee,
  Thiagarajan, Hersen, Chew, Safran, Zaidel-Bar, and Bershadsky}}]{Hu:2017}
\bibinfo{author}{\bibfnamefont{S.}~\bibnamefont{Hu}},
  \bibinfo{author}{\bibfnamefont{K.}~\bibnamefont{Dasbiswas}},
  \bibinfo{author}{\bibfnamefont{Z.}~\bibnamefont{Guo}},
  \bibinfo{author}{\bibfnamefont{Y.-H.} \bibnamefont{Tee}},
  \bibinfo{author}{\bibfnamefont{V.}~\bibnamefont{Thiagarajan}},
  \bibinfo{author}{\bibfnamefont{P.}~\bibnamefont{Hersen}},
  \bibinfo{author}{\bibfnamefont{T.-L.} \bibnamefont{Chew}},
  \bibinfo{author}{\bibfnamefont{S.~A.} \bibnamefont{Safran}},
  \bibinfo{author}{\bibfnamefont{R.}~\bibnamefont{Zaidel-Bar}},
  \bibnamefont{and} \bibinfo{author}{\bibfnamefont{A.~D.}
  \bibnamefont{Bershadsky}}, \bibinfo{journal}{Nat. Cell Biol.}
  \textbf{\bibinfo{volume}{19}}, \bibinfo{pages}{133} (\bibinfo{year}{2017}).

\end{thebibliography}
\end{document}